\documentclass[article, sort&compress]{elsarticle}

\usepackage{hyperref}
\usepackage{graphicx}
\usepackage{amsmath}
\usepackage{amssymb}
\usepackage{color}
\usepackage[normalem]{ulem}
\usepackage{lineno}

\makeatletter
\def\ps@pprintTitle{%
  \let\@oddhead\@empty
  \let\@evenhead\@empty
  \def\@oddfoot{\reset@font\hfil\thepage\hfil}
  \let\@evenfoot\@oddfoot
}
\makeatother

\begin{document}

\begin{frontmatter}

\title{Noise-induced, ac-stabilized sine-Gordon breathers: Emergence and statistics}


\author[unipa]{Duilio De Santis\corref{correspondingauthor}}
\cortext[correspondingauthor]{Corresponding author}
\ead{duilio.desantis@unipa.it}
\author[unisa,INFN]{Claudio Guarcello}
\author[unipa,unn]{Bernardo Spagnolo}
%\cortext[correspondingauthor]{Corresponding author}
%\ead{bernardo.spagnolo@unipa.it}
%
\author[unipa]{Angelo Carollo}
\author[unipa]{Davide Valenti}
\address[unipa]{Dipartimento di Fisica e Chimica ``E.~Segr\`{e}", Group of Interdisciplinary Theoretical Physics, Università degli Studi di 
Palermo, I-90128 Palermo, Italy}
\address[unisa]{Dipartimento di Fisica ``E.~R.~Caianiello", Università degli Studi di Salerno, I-84084 Fisciano, Salerno, Italy}
\address[INFN]{INFN, Sezione di Napoli, Gruppo Collegato di Salerno - Complesso Universitario di Monte
S. Angelo, I-80126 Napoli, Italy}
\address[unn]{Radiophysics Department, Lobachevsky State University, 603950 Nizhniy Novgorod, Russia}

\begin{abstract}
Noisy and ac forcing can cooperatively lead to the emergence of sine-Gordon breathers robust to dissipation. This phenomenon is studied, for both Neumann and periodic boundary conditions (NBC and PBC, respectively), at different values of the main system parameters, such as the noise intensity and the ac frequency-amplitude pair. In all the considered cases, nonmonotonicities of the probability of generating only breathers versus the noise strength are observed, implying that optimal noise ranges for the breather formation process exist. Within the latter scenarios, the statistics of the breathers' number, position, and amplitude are analyzed. The number of breathers is found to grow, on average, with the noise amplitude. The breathers' spatial distribution is sharply peaked at the system's edges for NBC, whereas it is essentially uniform for PBC. The average breather amplitude is dictated by the ac frequency-amplitude pair. Finally, a size analysis shows that the minimum system length for the generation mechanism is given by the typical breather half-width (width) in NBC (PBC).
\end{abstract}


\end{frontmatter}

%\linenumbers \modulolinenumbers[1]

\section{Introduction}
\label{Sec1}

The soliton is a cornerstone of nonlinear science, and it is nowadays ubiquitous in physics and beyond~\cite{Scott_2004, Scott_2006}. Two core reasons behind the widespread success of this concept are: (i)~the soliton displays remarkable particle-like properties which are both very robust to perturbations and compelling for a plethora of applications; (ii)~solitons mathematically emerge as solutions of partial differential equations involving fundamental types of nonlinearity that can be adopted for the description of many kinds of phenomena~\cite{Ablowitz_2000, Belinski_2001, Scott_2003, Dauxois_2006, Malomed_2022}. One of the most celebrated playgrounds for soliton dynamics is the one-dimensional sine-Gordon~(SG) equation, which constitutes a topic of never-ending interest for a broad spectrum of researchers~\cite{Jesus_2014}. Such a model, whose mechanical equivalent is a chain of pendula linked with linear springs~\cite{Scott_1969}, is indeed encountered in all sorts of scientific domains, including fundamental and applied superconductivity, crystal dislocations, ultra-short optical pulses, elementary particles, biology, geology, and gravity~\cite{Barone_1971, Ivancevic_2013, Bykov_2014, Jesus_2014}.

In the SG landscape, two basic building blocks of the solitonic spectrum can be identified: kinks/antikinks and breathers~\cite{Scott_2003, Dauxois_2006}. Kinks (antikinks) are step-like structures with positive (negative) topological charge. Interactions among pairs of kinks are possible and, understandably, they are repulsive in the kink-kink case and attractive in the kink-antikink one. The latter fact is responsible for the emergence of topologically neutral bound states, known as breathers, between kinks and antikinks. Breathers are exponentially localized in space and they `breathe' (i.e., oscillate) in time. Both theoretical and experimental issues related to the latter excitations, which represent the main concern of the below discussions, are of great importance in a significant number of scientific works. In particular, extensive investigations have been performed on SG breathers in various perturbed frameworks~\cite{Lomdahl_1986, Jensen_1992, Quintero_1998, Forinash_2001, Tao_2003, Jimenez_2004, Johnson_2013, Ferre_2017, Jagtap_2017, Raju_2021, De_Santis_2022_NES}, interacting-breather scenarios~\cite{Kevrekidis_2001, Nishida_2009, Jia_2020, Jia_2021}, biological systems~\cite{Ivancevic_2013, Liu_2021}, cuprate superconductors~\cite{Dienst_2013, Laplace_2016}, geological media~\cite{Bykov_2016, Zalohar_2020}, graphene-based structures~\cite{Belonenko_2009}, magnetic systems~\cite{Kiselev_2019}, liquid crystals~\cite{Li_2018, Li_2019}, network models~\cite{Caputo_2014, Dutykh_2018}, self-induced transparency media~\cite{Hou_2020}, and long Josephson junctions~(LJJs)~\cite{Ali_2011, Gulevich_2012, Gul_2018, De_Santis_2022, De_Santis_2022_CNSNS, De_Santis_2023, De_Santis_2023_TB}. Note that SG breathers are also found in two dimensions~\cite{Tamga_1995, Caputo_2013, Kevrekidis_2021} and that long-lived breather excitations can exist even in modified SG models~\cite{Ferreira_2016, Vergara_2021, Vergara_2022, Badikova_2023}.
% hydrogen storage materials~\cite{Hudak_2020}

Here, a study on the phenomenon of breather formation is presented in the noisy and ac-driven SG system with loss. Stochastic and sinusoidal forcing can indeed cooperatively lead to the emergence of long-time stable breathers in random locations~\cite{De_Santis_2023}. The manifestation of such events is investigated, for both Neumann and periodic boundary conditions (NBC and PBC, respectively), at different values of the main system parameters, such as the noise strength and the ac frequency-amplitude combination. Interestingly, nonmonotonic behaviors of the probability of generating exclusively breathers are obtained as a function of the noise intensity. This fact highlights the existence of optimal noise strength ranges for the above mechanism to occur. Once the most favorable scenarios for exciting breathers are identified, the statistics of their number, position, and amplitude are analyzed, and the following overall features are observed. The number of breathers grows, on average, with the noise strength. The breathers' spatial distribution displays sharp peaks at the system's edges for NBC, whereas it is essentially uniform for PBC. This qualitative difference is due to the existence of `edge-breather' modes~\cite{Costabile_1978, Cirillo_1981, Jensen_1992}, peculiar to the former boundary condition (BC) class, providing an energetic advantage with respect to their `bulk' counterparts. The average amplitude of the breather oscillations is dictated by the ac frequency-amplitude combination. More specifically, the typical breather amplitude is not quite that predicted by the unperturbed SG theory, but its value is readily understood within the damped-driven scenario. Lastly, the effect of the system size on the generation mechanism is analyzed, and a minimal length for its occurrence is found. This `critical' length is roughly given by the characteristic breather half-width (width) in NBC (PBC). The response's sensitivity to the BC type is, once again, explained by edge-breathers, which require half of the space than bulk excitations.

Direct applications of these findings include, but are not limited to, the notorious open problem of the breather's experimental observation in LJJs~\cite{Gulevich_2012, De_Santis_2022, De_Santis_2022_CNSNS, De_Santis_2023, De_Santis_2023_TB}. The considered generation technique was indeed recently employed for devising two detection schemes: (i)~a `destructive' protocol in which the breather's properties are embedded into the resistive switching statistics of the device~\cite{De_Santis_2023}; (ii)~a `non-destructive' approach exploiting the fact that a breather enhances the heat transport in a thermally biased LJJ~\cite{De_Santis_2023_TB}.

This work is structured as follows. Section~\ref{Sec2} describes the SG model, along with both its fundamental solitonic solutions in the unperturbed case and its physical interpretation in the context of LJJs. The results are illustrated in Sec.~\ref{Sec3} and, finally, conclusions are drawn in Sec.~\ref{Sec4}.

\section{Materials and methods}
\label{Sec2}

The analysis is concerned with the dissipative, harmonically-driven SG system in the presence of thermal noise. In dimensionless units, the equation for the field ${ \varphi (x, t) }$ reads~\cite{Scott_2003, Dauxois_2006}
\begin{equation}
\label{eqn:1}
\varphi_{xx} - \varphi_{tt} - \alpha \varphi_{t} = \sin \varphi - \eta \sin (\omega t) - \gamma_T (x, t) , \; \; x, t \in \left[ -\frac{l}{2}, \frac{l}{2} \right] \times \left[ 0, t_{\rm{sim}} \right] .
\end{equation}
Here, the notation ${ \partial \varphi / \partial x = \varphi_x }$ is adopted for the partial derivatives, ${ \alpha }$ is the dissipation coefficient, ${ \omega }$ and ${ \eta }$ are, respectively, the frequency and the amplitude of the external force, ${ \gamma_{T} (x, t) }$ is the zero-average Gaussian noise source, delta-correlated both in ${ x }$ and ${ t }$, with intensity ${ 2 \alpha \Gamma }$, ${ l }$ is the system length, and ${ t_{\rm{sim}} }$ is the simulation time.

Flat initial conditions are considered throughout the work for Eq.~\eqref{eqn:1} 
\begin{equation}
\label{eqn:2}
\varphi (x, 0) = \varphi_t (x, 0) = 0 ,
\end{equation}
along with either NBC
\begin{equation}
\label{eqn:3}
\varphi_x (-l/2, t) = \varphi_x (l/2, t) = 0 
\end{equation}
or PBC
\begin{equation}
\label{eqn:4}
\varphi (-l/2, t) = \varphi (l/2, t) .
\end{equation}

To properly set the stage for the discussions below, it is perhaps useful recalling that the perturbation-free version of Eq.~\eqref{eqn:1} (i.e., with ${ \alpha = \eta = \Gamma = 0 }$) is satisfied, for ${ l = \infty }$, by the elementary topological solitons~\cite{Scott_2003, Dauxois_2006}
\begin{equation}
\label{eqn:5}
\varphi_{\pm} (x, t) = 4 \arctan \left[ \exp \left( \pm \frac{x - vt}{\sqrt{1 - v^2}} \right) \right]
\end{equation} 
commonly labelled as kinks (${ \varphi_+ }$) and antikinks (${ \varphi_- }$). The nonlinear waves described by Eq.~\eqref{eqn:5} travel with constant velocity ${ v < 1 }$, and also display relativistic behavior. Within this context, the breather arises as a localized, oscillating bound state between a kink and an antikink, and its waveform~\cite{Scott_2003, Dauxois_2006}
\begin{equation}
\label{eqn:6}
\varphi_{b} (x, t) = 4 \arctan \left\lbrace \frac{\sqrt{1 - \omega_b^2}}{\omega_b} \frac{ \sin \left[ \frac{\omega_b \left( t - v_e x \right)}{\sqrt{1 - v_e^2}} \right] } { \cosh \left[ \frac{\sqrt{1 - \omega_b^2} \left( x - v_e t \right)}{\sqrt{1 - v_e^2}} \right] } \right\rbrace
\end{equation}
can indeed be understood as the analytic continuation of the profile that features a colliding kink-antikink pair. In Eq.~\eqref{eqn:6}, ${ 0 < \omega_b < 1 }$ and ${ v_e < 1 }$ are, respectively, the excitation's proper frequency and the envelope velocity. Note that the parameter ${ \omega_b }$ yields, for ${ v_e = 0 }$, the oscillation period ${ T_b = 2 \pi / \omega_b }$, the energy ${ E_b = 16 \sqrt{ 1 - \omega_b^2 } }$~\footnote[1]{The energy is obtained by integrating, over the whole spatial domain, the energy density ${
\varepsilon (x, t) = (\varphi_t^2 + \varphi_x^2)/2 + 1 - \cos \varphi }$ associated to the unperturbed SG equation, for ${ \varphi = \varphi_{b} }$.}, the amplitude ${ A_b = 4 \arctan \left( \sqrt{1 - \omega_b^2} / \omega_b \right) }$, and the characteristic length ${ \lambda_b \sim 1 / \sqrt{1 - \omega_b^2} }$ of the nonlinear mode. Thus, low (high) frequencies and high (low) oscillation amplitudes correspond to high (low) energy breathers. 

\begin{figure}[t!!]
\centering
\includegraphics[width=\textwidth]{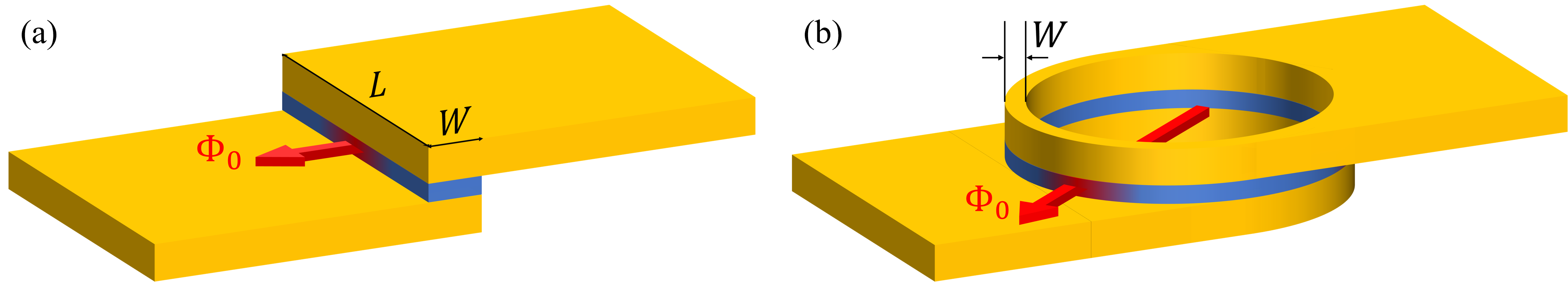}
\caption{Pictorial view of the LJJ geometries considered in the work. Panel~(a), i.e., the overlap case, corresponds to NBC [Eq.~\eqref{eqn:3}], while panel~(b), i.e., the annular case, to PBC [Eq.~\eqref{eqn:4}]. Single magnetic flux quanta ${ \Phi_0 }$, mathematically described by kinks, are sketched as red arrows. The symbols ${ L }$ and ${ W }$ indicate, respectively, the length and the width of the device.}
\label{fig:1}
\end{figure}

As mentioned in Sec.~\ref{Sec1}, the SG framework is successfully employed in a wide range of scientific contexts. The LJJ stands out among the latter as a prototypal physical system in condensed matter for exploring the rich SG phenomenology. In this regard, Eq.~\eqref{eqn:1} accounts for quasiparticle dissipation, a spatially-uniform alternating (ac) current, and thermal fluctuations, and it accurately reproduces the behavior of the driven superconducting device. The field ${ \varphi (x, t) }$ here represents the Josephson phase, i.e., the phase difference between the macroscopic wave functions relative to the two superconductors. The characteristic scales are given by the Josephson penetration depth ${ \lambda_J = \sqrt{ \Phi_0 / \left( 2 \pi J_c L_P \right)} }$ (${ l = L / \lambda_J }$ is the normalized junction length) and the Josephson plasma frequency ${ \omega_p = \sqrt{ 2 \pi J_c / \left( \Phi_0 C \right) } }$~\cite{Barone_1982, Lomdahl_1982}, where ${ \Phi_0 }$ is the magnetic flux quantum, ${ J_c }$ is the critical current per unit length, ${ L_P }$ is the inductance per unit length, and ${ C }$ is the capacitance per unit length. The dissipation constant reads ${ \alpha = G / \left( \omega_p C \right) }$ ($ G $ is the effective normal conductance), ${ \omega }$ (${ \eta }$) is given in units of ${ \omega_p }$ ($ J_c $), and the noise strength is ${ \Gamma = 2 e k_B T / \left( \hbar J_c \lambda_J \right) }$, in which ${ e }$ is the electron charge, ${ k_B }$ is the Boltzmann constant, ${ T }$ is the absolute temperature, and ${ \hbar }$ is the reduced Planck constant. Physically, Eqs.~\eqref{eqn:5}-\eqref{eqn:6} (more precisely, their perturbed counterparts) describe magnetic energy packets in the junction~\cite{Barone_1982, Lomdahl_1982}, and thus the above initial condition [Eq.~\eqref{eqn:2}] corresponds to an excitation-free initial state, whereas the BC specifies the device's geometry~\cite{Jesus_2014}: overlap for NBC [Eq.~\eqref{eqn:3}] and annular for PBC [Eq.~\eqref{eqn:4}], see Fig.~\ref{fig:1}. Various LJJ geometries have been studied over the years~\cite{Ustinov_1998, Wallraff_2003, Jesus_2014}, and the two considered here are, arguably, the most popular.

Equation~\eqref{eqn:1} is numerically solved by means of an implicit finite-difference algorithm, see Refs.~\cite{De_Santis_2022_NES, De_Santis_2023} for a detailed account on the matter. Throughout the paper, the discretization steps are ${ \Delta x = \Delta t = 0.01 }$, the system length is ${ l = 50 }$ unless otherwise specified, the simulation time is ${ t_{\rm{sim}} = 500 }$, the damping parameter is ${ \alpha = 0.2 }$~\cite{De_Santis_2023}, and the forcing frequency is below-plasma, i.e., ${ \omega < 1 }$, see Eq.~\eqref{eqn:6}.

\section{Results and discussion}
\label{Sec3}

\paragraph{Noise-induced, ac-locked breathers} As first discussed in Ref.~\cite{De_Santis_2023} for the PBC case, the combined action of thermal fluctuations and ac forcing can lead to the emergence, in random locations, of breathers surviving way beyond their radiative decay lifetime of ${ \sim 1 / \alpha = 5 }$~\cite{De_Santis_2022_NES}. More specifically, the excited modes are resonant with the external current source, their morphology closely resembles that of analytical breathers, e.g., frequency-dependent width and amplitude are exhibited in analogy with Eq.~\eqref{eqn:6}, and they also present a noteworthy stability with respect to position. Various questions on the statistics underlying this interesting phenomenon naturally arise and have yet to be quantitatively addressed: at given ${ \omega }$, ${ \eta }$, ${ \Gamma }$, and BC type, how likely is one to observe ${ n }$ breathers in a particular realization? How likely is a breather to be found at location ${ x }$? What is the average breather amplitude? Do any of these features depend on the noise strength ${ \Gamma }$? Moreover, Ref.~\cite{De_Santis_2023} mainly focuses on a single ac frequency-amplitude combination and PBC. Thus one may wonder how the latter choices influence the overall picture. A better understanding of these topics can surely be useful, especially in view of the long-sought experimental breather detection in LJJs, see Sec.~\ref{Sec1}.

Ideally, one would address the above queries by exploring a comprehensive set of physical scenarios, but this is rather impractical from a computational viewpoint. For each ac frequency-amplitude pair and BC type, a large enough set of runs must indeed be performed at different noise intensities for the results to be statistically relevant. Equation~\eqref{eqn:1} also has to be simulated for a sufficiently long time to let the breather generation events to unfold. In view of the previous facts, the choice of focusing on the medium-to-high breather frequency range is made by taking the combinations ${ \omega = 0.6 }$, ${ \eta = 0.59 }$ and ${ \omega = 0.7 }$, ${ \eta = 0.44 }$~\footnote[2]{The amplitude ${ \eta }$ paired to each ${ \omega }$ value is the lowest at which the phenomenon of interest is observed. The exclusive generation of breathers is usually possible for a range of ac amplitudes above the analyzed ones, see Fig.~3 in Ref.~\cite{De_Santis_2023}, but it is reasonable to first consider the cases closer to the unperturbed SG limit.}. For those frequency-amplitude pairs, both in NBC and PBC, ensembles of ${ N = 5000 }$ numerical experiments are examined at different ${ \Gamma }$ values.

\begin{figure}[t!!]
\centering
\includegraphics[width=0.85\textwidth]{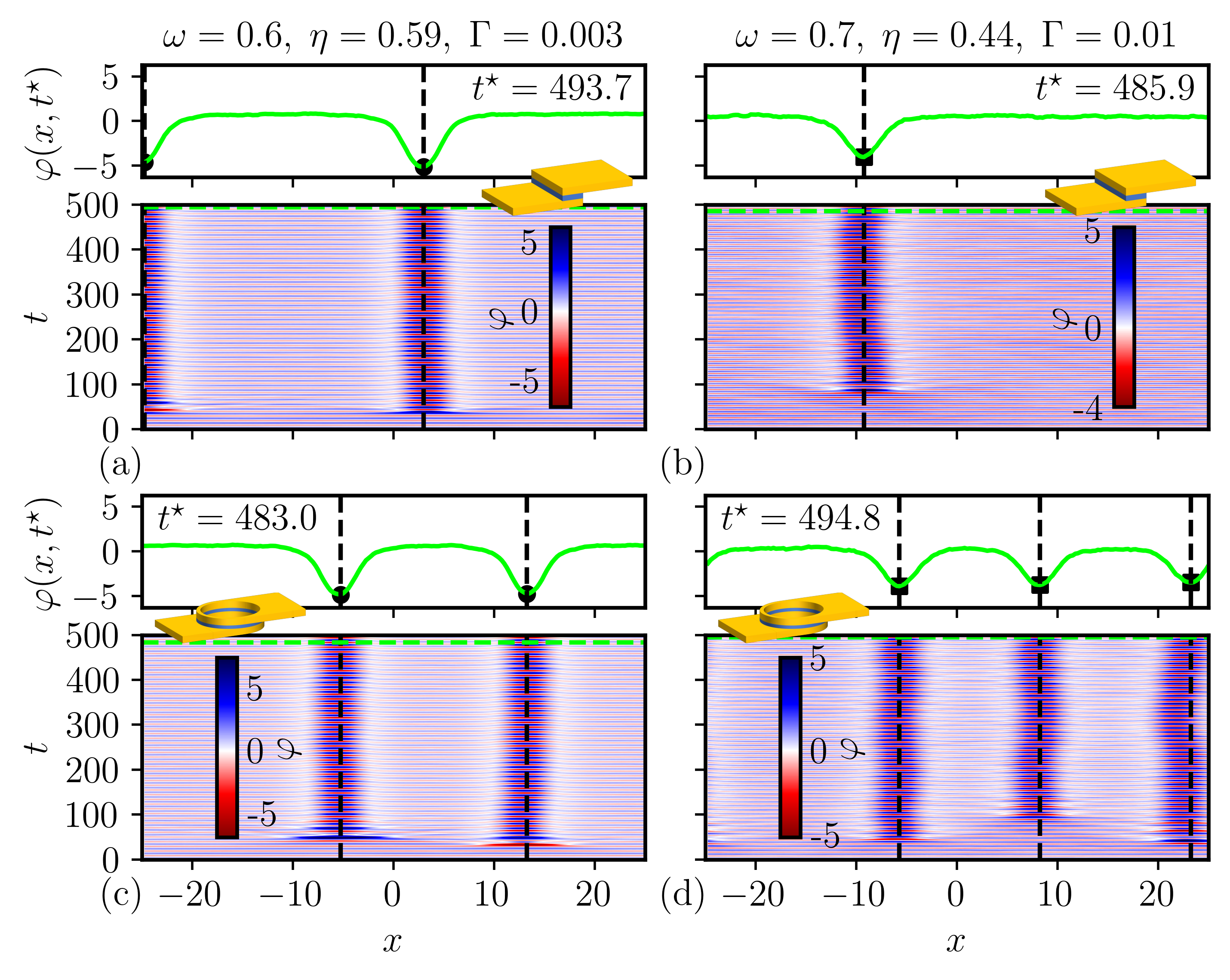}
\caption{Four different numerical realizations. In each panel, the main plot shows the spatio-temporal evolution of ${ \varphi }$, whereas the inset corresponds to the phase snapshot ${ \varphi (x, t^\star) }$ (the instant ${ t = t^\star }$ is indicated with a dashed green horizontal line in the contour graph), which is used to analyze the number, position, and amplitude of the excited breathers, see the circles/squares and dashed vertical lines in black. The results on the left are obtained with ${ \omega = 0.6 }$, ${ \eta = 0.59 }$, and ${ \Gamma = 0.003 }$, whereas those on the right with ${ \omega = 0.7 }$, ${ \eta = 0.44 }$, and ${ \Gamma = 0.01 }$. Furthermore, each panel features a sketch of the LJJ geometry relative to the BC imposed in the simulation. Shared parameter values: ${ \Delta x = \Delta t = 0.01 }$, ${ l = 50 }$, ${ t_{\rm{sim}} = 500 }$, and ${ \alpha = 0.2 }$.}
\label{fig:2}
\end{figure}
%
% main vs inset plot
% oppure contour vs line plot
To operatively illustrate how the analysis works, it is worth starting from some examples. Figure~\ref{fig:2} displays four runs: the two on the left are obtained with ${ \omega = 0.6 }$, ${ \eta = 0.59 }$, and ${ \Gamma = 0.003 }$, whereas those on the right with ${ \omega = 0.7 }$, ${ \eta = 0.44 }$, and ${ \Gamma = 0.01 }$. Each panel also features a sketch of the LJJ geometry relative to the BC imposed in that specific simulation. The distinctive traits of the phenomenon, summarized above, can be identified in all the contour plots, which present the spatio-temporal evolution of the phase ${ \varphi }$. In particular, one or more breathers are seen to emerge in random spots within a few driving cycles, and they persist up to ${ t_{\rm{sim}} = 500 }$~\footnote[3]{As one appreciates from Fig.~1 in Ref.~\cite{De_Santis_2023}, the nonlinear modes can survive even for ${ t > 500 }$.}. A peculiar feature of NBC, already evident in Fig.~\ref{fig:2}(a), is that single kinks/antikinks can form bound states with virtual companions at the boundaries. These edge-breathers, which can also be encountered in the earlier SG literature~\cite{Costabile_1978, Cirillo_1981, Jensen_1992}, are intuitively expected to be more likely than the bulk counterparts due to their reduced energy cost.

The focus of the present study is exclusively on the cases where, similarly to Fig.~\ref{fig:2}(a)-(c), breathers are the only emerging excitations. The devised strategy for analyzing these runs is to focus on a suitable instant ${ t = t^\star \lesssim t_{\rm{sim}} }$ and apply a peak detection algorithm~\footnote[4]{\texttt{Scipy}'s \texttt{signal.find{\textunderscore}peaks} is employed here. Via extensitve preliminary testing, the routine's \texttt{prominence} parameter is tuned to capture breather-like modes very reliably.} to the profile ${ \varphi (x, t^\star) }$, see the insets in Fig.~\ref{fig:2}. Such an approach takes advantage of the fact that the excited breathers share the same oscillation cycle, locked to the ac force, in every simulation. If this were not true, it clearly would not be possible, in general, to measure all the breather amplitudes [see, e.g., the black circles/squares in Fig.~\ref{fig:2}] through a single ${ \varphi }$ snapshot. Moreover, an underlying assumption is that the configuration at ${ t^\star \lesssim t_{\rm{sim}} }$ is somewhat representative of the realization. This is fairly true, especially for the moderate noise amplitude values considered for the statistical analysis. The observed features are indeed mostly unaltered, after the occurrence of the generation events over a rather short transient, for hundreds of breathing periods. For example, the black dashed vertical lines in Fig.~\ref{fig:2} indicate the detected breather positions at ${ t = t^\star }$, and they are consistent with the entire simulated evolution. One last remark concerns the time ${ t^\star }$, which is operatively chosen, within the final two driving cycles of each run, as the instant characterized by the maximal phase excursion, see the dashed green horizontal lines in the contour plots in Fig.~\ref{fig:2}.

\begin{figure}[t!!]
\centering
\includegraphics[width=0.8\textwidth]{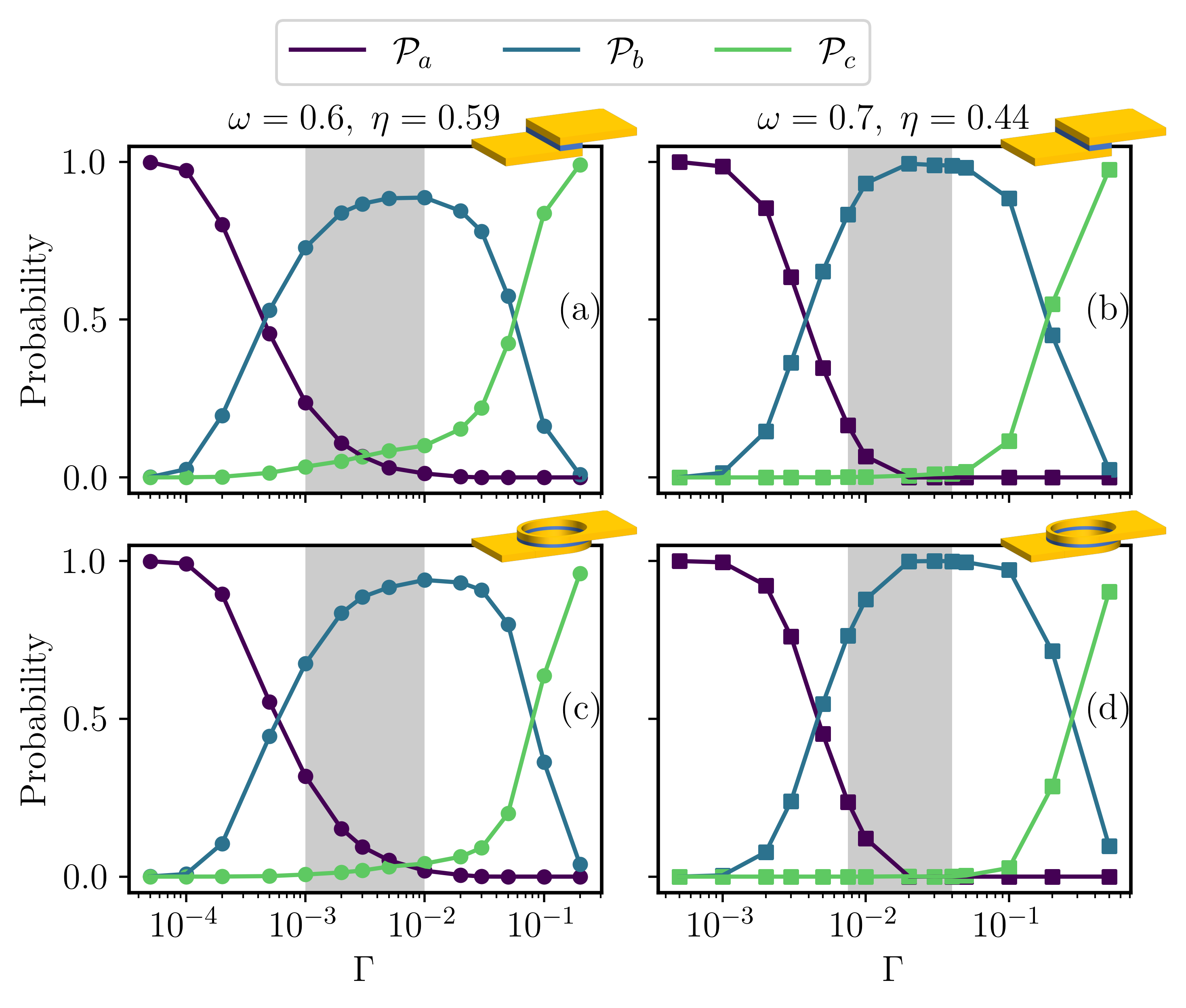}
\caption{Probabilities of having no excitations (${ \mathcal{P}_a }$, purple), breathers only (${ \mathcal{P}_b }$, blue), and at least one kink-type excitation (${ \mathcal{P}_c }$, green) versus ${ \Gamma }$. The results on the left (circles) are obtained with ${ \omega = 0.6 }$ and ${ \eta = 0.59 }$, whereas those on the right (squares) with ${ \omega = 0.7 }$ and ${ \eta = 0.44 }$. Each panel features a sketch of the LJJ geometry relative to the BC imposed in the simulations. Also, the gray shaded areas denote the ranges of noise strengths selected for the statistical analysis: ${ \Gamma \in [0.001, 0.01] }$ for panels (a) and (c); ${ \Gamma \in [0.007, 0.04] }$ for panels (b) and (d). Shared parameter values: ${ \Delta x = \Delta t = 0.01 }$, ${ l = 50 }$, ${ t_{\rm{sim}} = 500 }$, ${ \alpha = 0.2 }$, and ${ N = 5000 }$.}
\label{fig:3}
\end{figure}
Reasonable ranges of the ${ \Gamma }$ parameter for the statistical analysis should involve a high fraction of `breathers only' cases. To this end, the probability ${ \mathcal{P}_b }$ of generating exclusively breathers is evaluated as a function of the noise strength, along with the probability of having no excitations ${ \mathcal{P}_a }$, and that of exciting at least one kink-type structure ${ \mathcal{P}_c }$. As illustrated in Fig.~\ref{fig:3}, whose left panels [(a) and (c), see the circles] are obtained with ${ \omega = 0.6 }$ and ${ \eta = 0.59 }$, whereas those on the right [(b) and (d), see the squares] with ${ \omega = 0.7 }$ and ${ \eta = 0.44 }$, the behavior is quite similar for NBC and PBC. For the smallest noise intensities, the probability ${ \mathcal{P}_a }$ of having no excitations is ${ 1 }$, see the purple curves. For intermediate ${ \Gamma }$ values, ${ \mathcal{P}_b }$ peaks at ${ \gtrsim 0.9 }$, see the blue curves. A further increase of the thermal fluctuations leads to less coherent breathers, and kink-type structures begin to take over, as indicated by the ${ \mathcal{P}_c }$ (green) curves. It may be worth observing that, in both the left and right panels of Fig.~\ref{fig:3}, for NBC the probability ${ \mathcal{P}_b }$ rises (falls) a bit faster than for PBC, at lower (higher) noise intensities, a fact which is likely due to the existence of the easier-to-create (easier-to-destroy) edge-breathers in the former BC type. Another relevant aspect of Fig.~\ref{fig:3} is that the ${ \mathcal{P}_b (\Gamma) }$ curves at ${ \omega = 0.6 }$ and ${ \eta = 0.59 }$, see the blue circles, reach their peak earlier than their counterparts at ${ \omega = 0.7 }$ and ${ \eta = 0.44 }$, see the blue squares. Based on these findings, the below investigation takes place in the gray shaded ${ \Gamma }$ regions in Fig.~\ref{fig:3}(a)-(c), which extend roughly over one order of magnitude around the ${ \mathcal{P}_b (\Gamma) }$ plateaus, leaving out the noise intensities associated with the transition towards the regime of significant thermal agitation (signaled by ${ \mathcal{P}_c }$'s growth). More quantitatively, for ${ \omega = 0.6 }$ and ${ \eta = 0.59 }$ (${ \omega = 0.7 }$ and ${ \eta = 0.44 }$) the highlighted area is ${ \Gamma \in [0.001, 0.01] }$ (${ \Gamma \in [0.007, 0.04] }$).

\begin{figure}[t!!]
\centering
\includegraphics[width=0.8\textwidth]{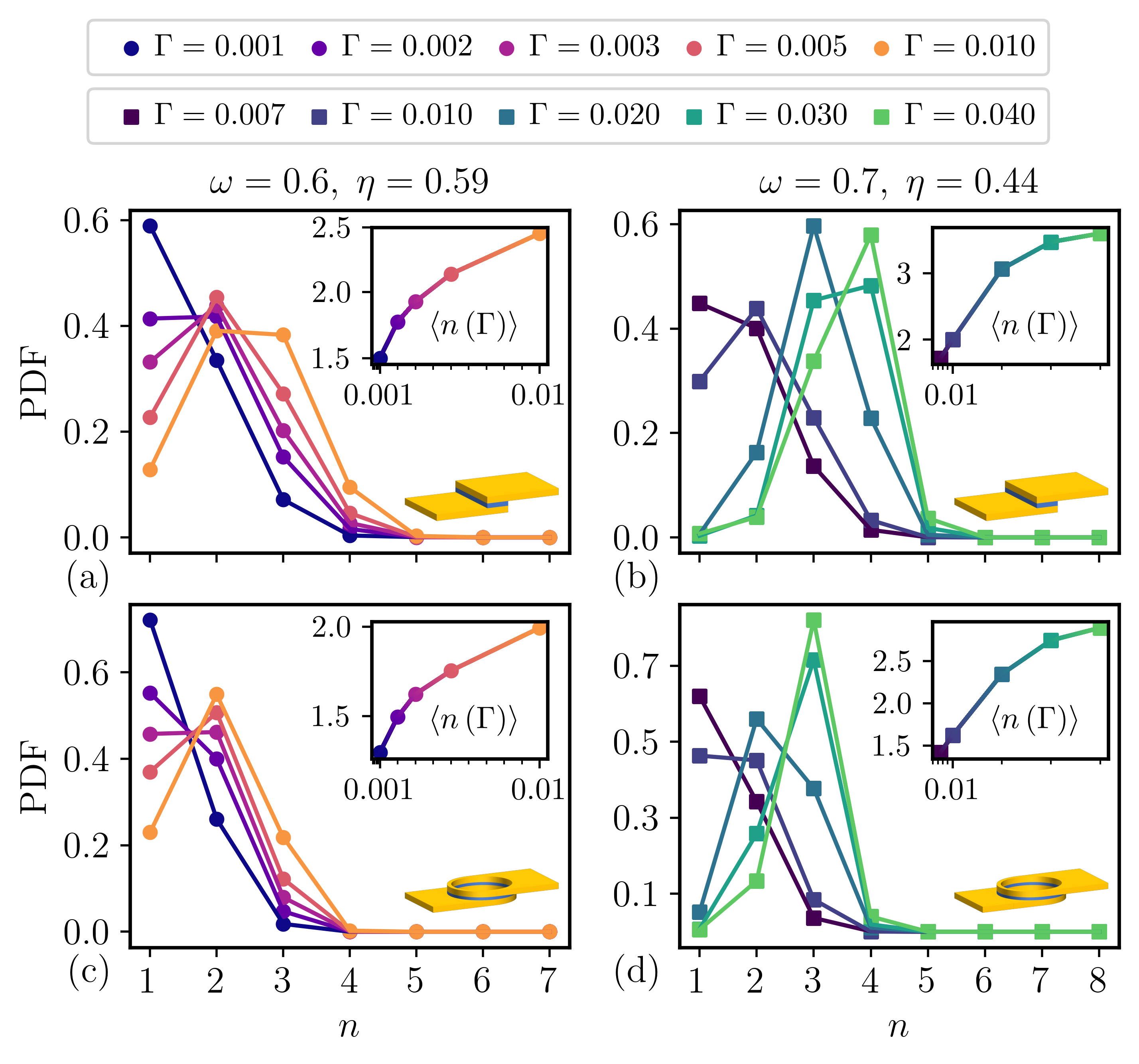}
\caption{\textit{Main plots:}~PDFs of the number of breathers ${ n }$ observed at different noise strengths. \textit{Insets:}~Averages of the breather number, ${ \left\langle n \right\rangle }$, as a function of ${ \Gamma }$ (which takes values in the same sets as in the corresponding main panels). The results on the left (circles) are obtained with ${ \omega = 0.6 }$ and ${ \eta = 0.59 }$, whereas those on the right (squares) with ${ \omega = 0.7 }$ and ${ \eta = 0.44 }$. Each panel also features a sketch of the LJJ geometry relative to the BC imposed in the simulations. Shared parameter values: ${ \Delta x = \Delta t = 0.01 }$, ${ l = 50 }$, ${ t_{\rm{sim}} = 500 }$, ${ \alpha = 0.2 }$, and ${ N = 5000 }$.}
\label{fig:4}
\end{figure}
\paragraph{Breather statistics} Figure~\ref{fig:4}'s main plots present the probability density functions~(PDFs) of the breather number ${ n }$ for different noise strengths, while the insets show the average breather numbers ${ \left\langle n (\Gamma) \right\rangle }$. Like before, panels (a) and (c) [(b) and (d)] are obtained with ${ \omega = 0.6 }$ and ${ \eta = 0.59 }$ [${ \omega = 0.7 }$ and ${ \eta = 0.44 }$]. Upon increase of the noise intensity, all the PDF peaks are seen to shift from ${ n = 1 }$ towards higher ${ n }$ values, i.e., it gradually becomes more probable to observe several breathers in each run, see the circles/squares. Generally speaking, PBC seem characterized by less breathers than NBC, as one appreciates by comparing panels (a) and (b) with panels (c) and (d), respectively. This information can also be deduced from the insets, where monotonically growing profiles of the average breather number are displayed. Overall, it should be clear from Fig.~\ref{fig:4} that the noise amplitude represents an useful control parameter, since it allows to tune the quantity ${ \left\langle n \right\rangle }$.

\begin{figure}[t!!]
\centering
\includegraphics[width=0.85\textwidth]{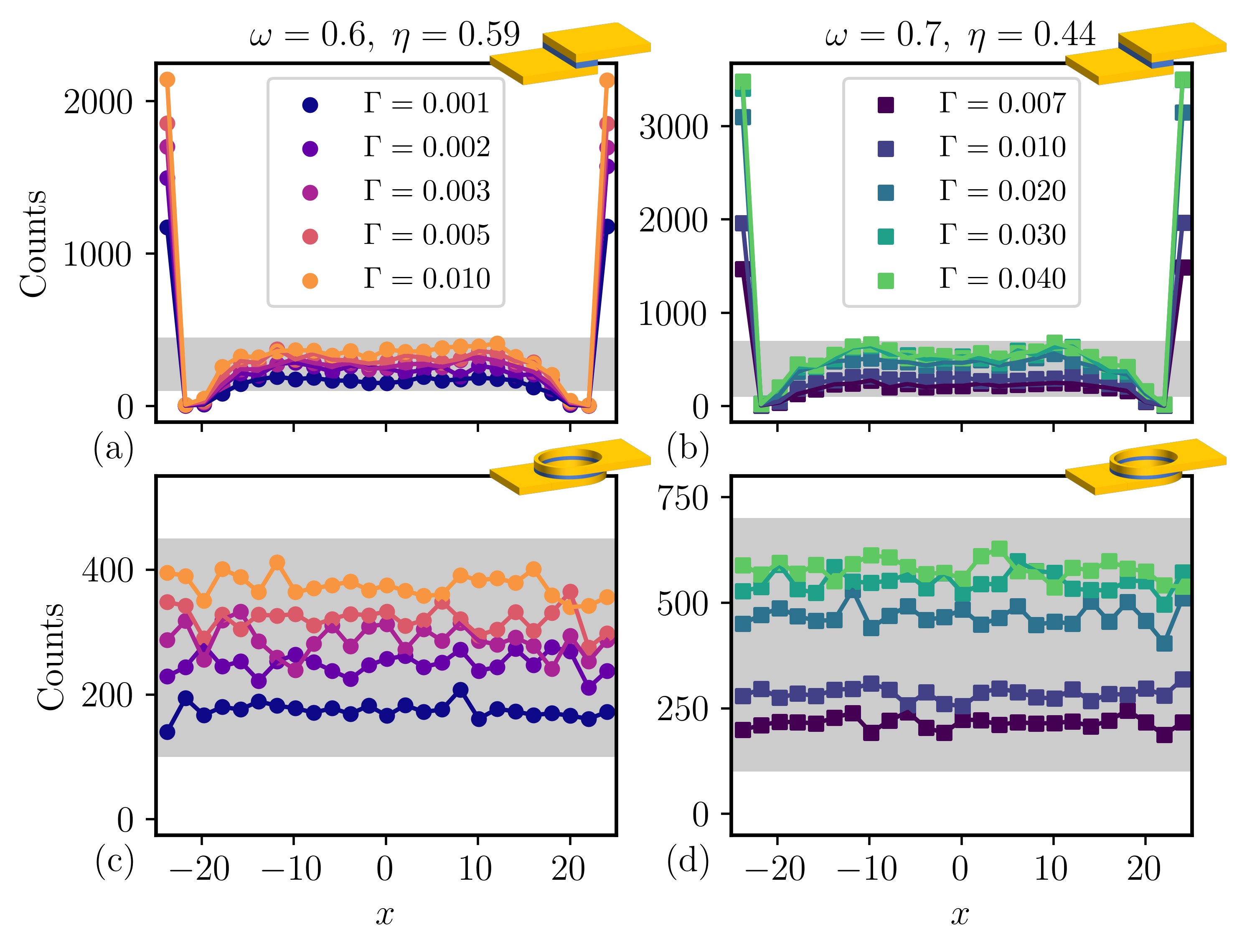}
\caption{Breather counts as a function of the position ${ x }$, for different noise strengths ${ \Gamma }$. The results on the left (circles) are obtained with ${ \omega = 0.6 }$ and ${ \eta = 0.59 }$, whereas those on the right (squares) with ${ \omega = 0.7 }$ and ${ \eta = 0.44 }$. The legends in panels (a) and (b) also apply to panels (c) and (d), respectively. The gray shaded areas indicate the ranges ${ [100, 450] }$ [panels (a) and (c)] and ${ [100, 700] }$ [panels (b) and (d)]. Furthermore, each plot features a sketch of the LJJ geometry relative to the BC imposed in the simulations. Shared parameter values: ${ \Delta x = \Delta t = 0.01 }$, ${ l = 50 }$, ${ t_{\rm{sim}} = 500 }$, ${ \alpha = 0.2 }$, and ${ N = 5000 }$.}
\label{fig:5}
\end{figure}
Another topic of interest, that is, the likelihood of a breather being found at a given position, is addressed in Fig.~\ref{fig:5}. More precisely, the plots illustrate the breather counts versus ${ x }$, for different noise strengths, with ${ \omega = 0.6 }$ and ${ \eta = 0.59 }$ (left panels, circles) and ${ \omega = 0.7 }$ and ${ \eta = 0.44 }$ (right panels, squares). For both frequency-amplitude pairs, a striking difference between the upper (NBC) and lower (PBC) results emerges. The NBC case indeed features distributions that are sharply peaked at the boundaries, whereas the PBC counterparts are essentially uniform. These outcomes are well understood: being essentially single-soliton modes, edge-breathers can be excited more easily for NBC than their bulk analogues, while the absence of preferred locations for the generation events is expected for PBC. Concerning the NBC panels, note also that, for ${ | x | \lesssim l/2 }$, regions with near-zero breather counts extend over the typical modes' width due to boundary interactions; on the other hand, a behavior similar to PBC is recovered for ${ | x | \ll l/2 }$, as seen by direct comparison with the visual aid of the gray shaded areas. Figure~\ref{fig:5} allows one to conclude that the BC type significatively impacts on the spatial distribution of breathers~\footnote[5]{Furthermore, Fig.~\ref{fig:5} explains the systematic difference in the averages of the breather number ${ n }$ between NBC and PBC, see Fig.~\ref{fig:4}.}. Instead, changing the value of the noise strength ${ \Gamma }$ seems to play a role mainly in shifting the distributions' average.

\begin{figure}[t!!]
\centering
\includegraphics[width=0.85\textwidth]{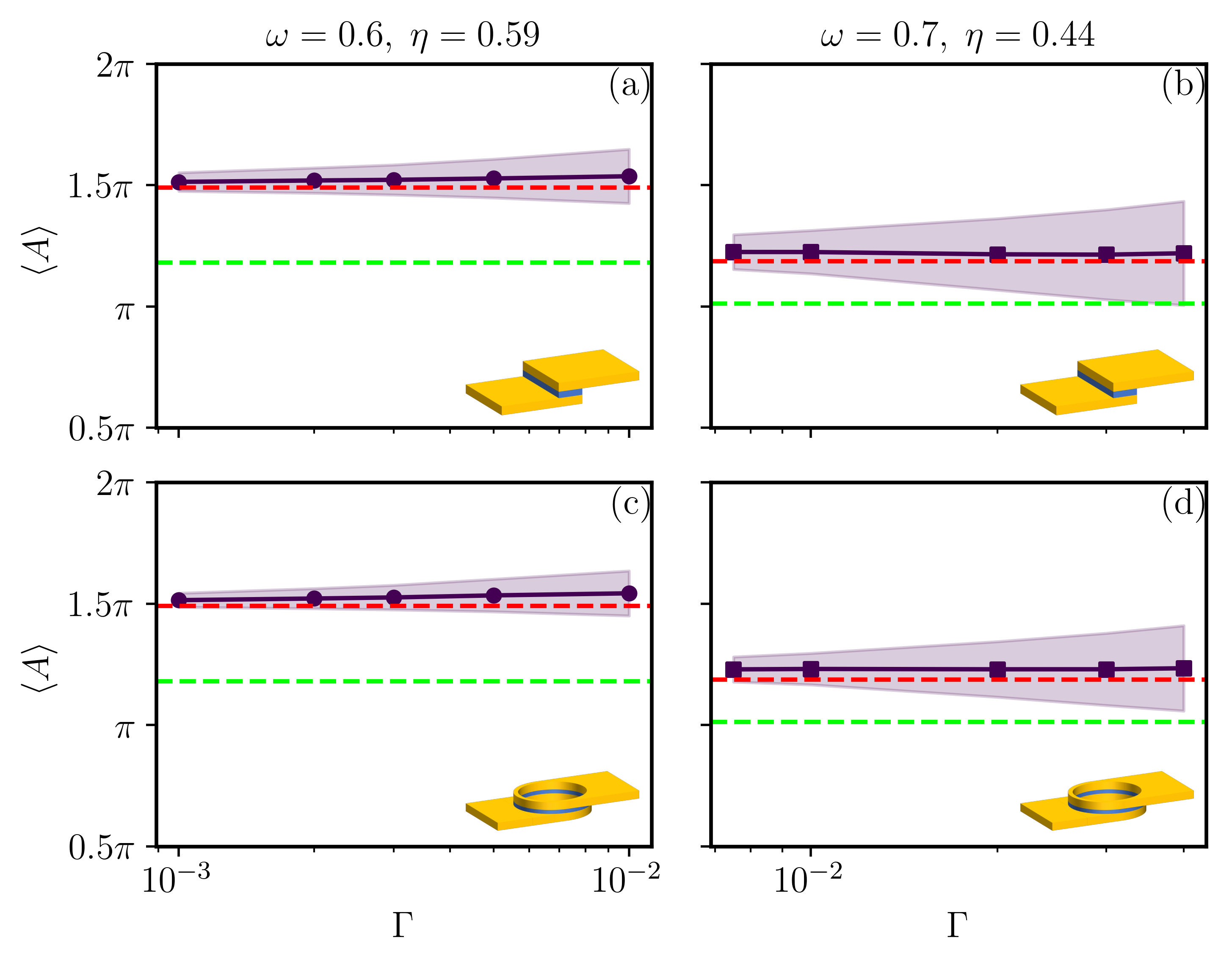}
\caption{Averages of the breather amplitude, ${ \left\langle A \right\rangle }$, and their standard deviations (shaded areas), as a function of ${ \Gamma }$. The results on the left (circles) are obtained with ${ \omega = 0.6 }$ and ${ \eta = 0.59 }$, whereas those on the right (squares) with ${ \omega = 0.7 }$ and ${ \eta = 0.44 }$. The dashed green horizontal lines denote the analytical breather amplitudes, see Eq.~\eqref{eqn:6}, at ${ \omega = 0.6 }$ [panels (a) and (c)] and at ${ \omega = 0.7 }$ [panels (b) and (d)]. The dashed red horizontal lines indicate, instead, the amplitudes observed by simulating the noiseless damped-driven system [${ \omega = 0.6 }$ and ${ \eta = 0.59 }$ for panels (a) and (c), ${ \omega = 0.7 }$ and ${ \eta = 0.44 }$ for panels (b) and (d)], with ad-hoc analytical breather initial conditions. Each plot also features a sketch of the LJJ geometry relative to the BC imposed in the numerical experiments. Shared parameter values: ${ \Delta x = \Delta t = 0.01 }$, ${ l = 50 }$, ${ t_{\rm{sim}} = 500 }$, ${ \alpha = 0.2 }$, and ${ N = 5000 }$.}
\label{fig:6}
\end{figure}
Consider now the averages of the breather amplitude ${ \left\langle A \right\rangle }$, which are plotted in Fig.~\ref{fig:6} (circles and squares), along with their standard deviations (shaded areas), as a function of the noise intensity. Here, the same qualitative trends are obtained in all cases. The ${ \left\langle A (\Gamma) \right\rangle }$ curves are indeed almost constant, and the corresponding standard deviations gradually increase with ${ \Gamma }$. The fact that panels (a) and (b) are practically identical, respectively, to panels (c) and (d) suggests that the BC are not very influential on the average breather amplitude. The latter quantity turns out to be dictated by the frequency-amplitude pair, as shown by the systematic difference between the simulation results on the left and those on the right~\footnote[6]{This feature can be appreciated also in the plots of Fig.~\ref{fig:2}, see the insets.}. Note that the observed ${ \left\langle A \right\rangle }$ values are higher than their unperturbed counterparts, i.e., the analytical breather amplitudes calculated via Eq.~\eqref{eqn:6} and denoted by the dashed green horizontal lines, see Fig.~\ref{fig:6}. A very close match for the ensemble averages is found with the amplitudes extracted from separate numerical experiments, with ad-hoc analytical breather initial conditions, in the noise-free (${ \Gamma = 0 }$) damped-driven scenario, see the dashed red horizontal lines. In other words, the typical amplitudes exhibited by noise-induced breathers are nearly equal to those at which exact breathers spontaneously relax in the presence of dissipative and ac perturbations [${ \omega = 0.6 }$ and ${ \eta = 0.59 }$ for panels (a) and (c), ${ \omega = 0.7 }$ and ${ \eta = 0.44 }$ for panels (b) and (d)]. Finally, in regard to the standard deviations displayed in Fig.~\ref{fig:6}, more scattered breather amplitude values constitute a reasonable outcome in the presence of a progressively larger amount of noise.

\paragraph{Size analysis} One last aspect that we wish to highlight is the effect of the system size ${ l }$ on the breathers' emergence mechanism. In particular, one might wonder whether a minimum junction length exists for the generation events to occur. To this end, parameter sets corresponding to peaks of the breather-only probability ${ \mathcal{P}_b }$, see Fig.~\ref{fig:3}, are now chosen: (i) ${ \omega = 0.6 }$, ${ \eta = 0.59 }$, and ${ \Gamma = 0.003 }$; (ii) ${ \omega = 0.7 }$, ${ \eta = 0.44 }$, and ${ \Gamma = 0.01 }$. For each of the latter, both in NBC and PBC, ensembles of ${ N = 5000 }$ runs are performed at different sizes ${ l \in \left[ 1, 32 \right] }$.

\begin{figure}[t!!]
\centering
\includegraphics[width=0.85\textwidth]{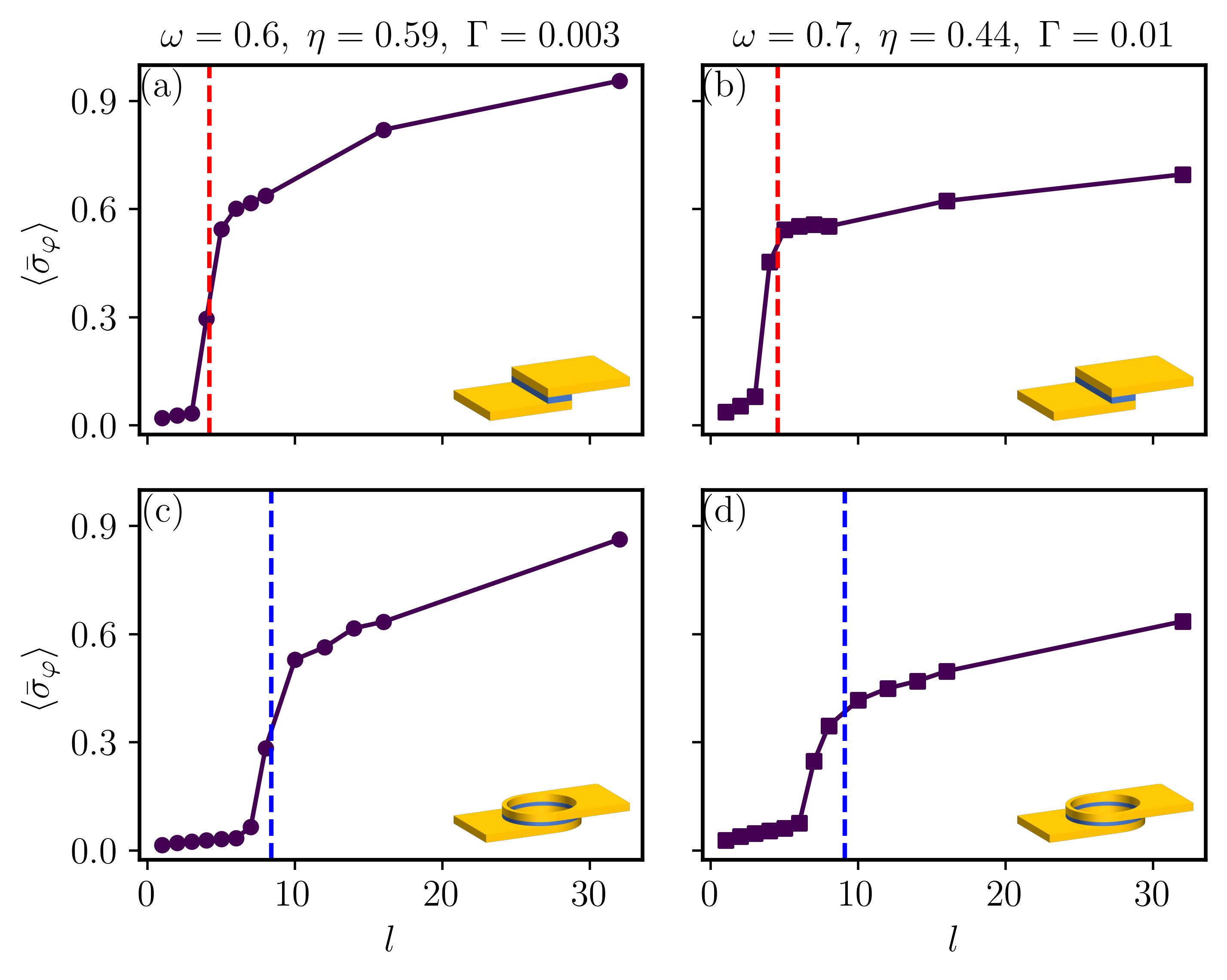}
\caption{Ensemble averages of the spatial standard deviation ${ \bar{\sigma}_{\varphi} }$, defined in Eq.~\eqref{eqn:7}, as a function of the size ${ l }$. The results on the left (circles) are obtained with ${ \omega = 0.6 }$, ${ \eta = 0.59 }$, and ${ \Gamma = 0.003 }$, whereas those on the right (squares) with ${ \omega = 0.7 }$, ${ \eta = 0.44 }$, and ${ \Gamma = 0.01 }$. The dashed red vertical lines denote the breather half-widths, computed via Eq.~\eqref{eqn:6}, at ${ \omega = 0.6 }$ [panel (a)] and at ${ \omega = 0.7 }$ [panel (b)]. The dashed blue vertical lines indicate the breather widths at ${ \omega = 0.6 }$ [panel (c)] and at ${ \omega = 0.7 }$ [panel (d)]. Each plot also features a sketch of the LJJ geometry relative to the BC imposed in the numerical experiments. Shared parameter values: ${ \Delta x = \Delta t = 0.01 }$, ${ t_{\rm{sim}} = 500 }$, ${ \alpha = 0.2 }$, ${ N = 5000 }$, and ${ t_0 = 250 }$.}
\label{fig:7}
\end{figure}
To efficiently characterize the response at different ${ l }$ values, the following time-averaged spatial standard deviation of the phase is introduced 
\begin{equation}
\label{eqn:7}
\bar{\sigma}_{\varphi} = \frac{1}{t_{\rm{sim}} - t_0} \int_{t_0}^{t_{\rm{sim}}} dt \; \sigma_\varphi (t) ,
\end{equation}
with ${ t_0 = 250 }$ being a fixed time instant~\footnote[7]{Different values have been extensively tested, but the results presented in Fig.~\ref{fig:7} show essentially no dependence on the choice of ${ t_0 < t_{\rm{sim}} }$ within reasonable limits.}, and ${ \sigma_\varphi (t) }$ the standard deviation of the instantaneous curve ${ \varphi (x, t) }$. The quantity considered in Eq.~\eqref{eqn:7} is intuitively near-zero in situations free of solitonic excitations, i.e., for phase profiles close to spatially flat. By contrast, whenever ${ \varphi }$ is inhomogeneous in space due to the presence of persistent breathers, ${ \bar{\sigma}_{\varphi} }$ assumes larger values.

\sloppy Ensemble averages of ${ \bar{\sigma}_{\varphi} }$, over ${ N = 5000 }$ simulations, are shown versus the size ${ l }$ in Fig.~\ref{fig:7}. The results on the left (circles) are obtained with ${ \omega = 0.6 }$, ${ \eta = 0.59 }$, and ${ \Gamma = 0.003 }$, while those on the right (squares) with ${ \omega = 0.7 }$, ${ \eta = 0.44 }$, and ${ \Gamma = 0.01 }$. Notably, all the ${ \left\langle \bar{\sigma}_{\varphi} (l) \right\rangle }$ curves display rather sharp transitions at precise system lengths. By comparing panels (a) and (b) with, respectively, panels (c) and (d), one immediately sees that the critical ${ l }$ value for NBC is roughly half of that in the corresponding PBC case. This behavior is ultimately due to the fact that a junction must be long enough to fit a breather for its emergence to be possible. For PBC the transition occurs when ${ l }$ is roughly equal to the characteristic breather width, which can be evaluated via Eq.~\eqref{eqn:6}, see dashed blue vertical lines in panels (c) and (d). On the other hand, edge-breathers are allowed in NBC, and these modes require half of the space compared to their bulk counterparts. The NBC critical length is then, quite reasonably, very close to the characteristic breather half-width, see the dashed red vertical lines in panels (a) and (b).

\section{Conclusions}
\label{Sec4}

This study is concerned with the emergence of breathers, for both NBC and PBC, in the damped and ac-driven SG equation in the presence of thermal noise. Different ac frequency-amplitude pairs are examined, and the probability of breather-only generation is seen to behave nonmonotonically versus the noise intensity. Within the most favorable scenarios for exciting breathers, the statistics of their number, position, and amplitude are addressed. In particular, increasing the noise strength makes the number of breathers grow on average. The breather distribution in space presents, for NBC, pronounced peaks at the boundaries, while, for PBC, it is approximately flat. The reason behind such a difference is the edge-breather mode, peculiar to NBC, that is easier to excite than its bulk counterpart. Away from the edges, a similar behavior for NBC and PBC is recovered, as one may imagine. The average amplitude at which the breather oscillations occur is determined by the ac frequency-amplitude pair, a fact that is well-understood in terms of the damped-driven SG scenario. Finally, the analysis looks at the influence of the system length on the overall picture. The existence of minimal length for generating breathers, which is roughly equal to the breather half-width (width) in NBC (PBC), is highlighted. Again, edge-breathers, which occupy half of the space compared to bulk modes, explain the results' sensitivity to the different BC.

The above findings are clearly useful for, e.g., devising an experimental configuration suitable for a breather detection scheme in LJJs~\cite{Gulevich_2012, De_Santis_2022, De_Santis_2022_CNSNS, De_Santis_2023, De_Santis_2023_TB}. The provided information on the effects of the system's main parameters, geometry, and size on the generation mechanism allows for tailoring a source of remarkably stable breathers. Quantities such as the average number of excited breathers are indeed shown to be controllable in conceptually simple ways, for example, by changing the noise strength.

\section*{Acknowledgments}
\label{acknowledgments}

The majority of the above numerical runs were performed on CINECA's machine Galileo100, Project: IscrC{\textunderscore}NDJB. The authors acknowledge the support of the Italian Ministry of University and Research (MUR).

%\bibliography{Long_NSGB_De_Santis_Refs}

\end{document}